\documentclass[12pt,preprint]{aastex}
\newcommand{\be}{\begin{equation}}
\newcommand{\ee}{\end{equation}}

\shorttitle{Line-of-sight dust at high $z$}
\shortauthors{Trenti \& Stiavelli}

\begin{document}

%% LaTeX will automatically break titles if they run longer than
%% one line. However, you may use \\ to force a line break if
%% you desire.

\title{Dust absorption along the line of sight for high-redshift
objects}

%% Use \author, \affil, and the \and command to format
%% author and affiliation information.
%% Note that \email has replaced the old \authoremail command
%% from AASTeX v4.0. 
%% You can use \email to mark an email address
%% anywhere i the paper, not just in the front matter.
%% As in the title, use \\ to force line breaks.

\author{M. Trenti \and M. Stiavelli} \affil{Space Telescope Science
Institute, 3700 San Martin Drive Baltimore MD 21218 USA}
\email{trenti@stsci.edu; mstiavel@stsci.edu}

%% Notice that each of these authors has alternate affiliations, which
%% are identified by the \altaffilmark after each name.  Specify alternate
%% affiliation information with \altaffiltext, with one command per each
%% affiliation.

\begin{abstract}

We estimate the optical depth distribution of dust present in
absorption systems along the line of sight of high redshift galaxies
and the resulting reddening. We characterize the probability
distribution of the transmission to a given redshift and the shape of
the effective mean extinction law by means of analytical estimates and
Monte Carlo simulations.  We present our results in a format useful
for applications to present samples of high redshift galaxies and
discuss the implications for observations with the \emph{James Webb
Space Telescope}. Our most realistic model takes into account the
metallicity evolution of Damped Lyman $\alpha$ absorbers and predicts
that the effects of dust absorption are modest: at redshift $z \gtrsim
5$ the transmission is above $0.8$ at an emitted wavelength $\lambda_e
= 0.14 \mu m$ with probability $90\%$. Therefore dust obscuration
along the line of sight will affect only marginally observations at
very high redshift.

\end{abstract}


%%%%%%%%%%%%%%%%%%%%%%%%%%%%%%%%%%%%
\section{Introduction}

Dust along the line of sights affects the observations of distant
astronomical objects and its effects on both extinction and reddening
need to be taken into account. In the case of high redshift quasars
this problem was initially addressed by \citet{ost84} and it has been
the subject of several investigations thereafter. \citet{fal89a,fal93}
developed a comprehensive theoretical framework to quantify the
effects of dust along the line of sight and characterize sample
selection effects induced by Damped Lyman $\alpha$ (hereafter DLA)
absorbers. The reddening measurements by \citet{pei91}, based on a
sample of 13 DLA systems with average absorption redshift $\langle
z_{abs} \rangle \approx 2.6$, evidenced a reddening of background
quasars with DLA spectral fingerprints with respect to a control
sample without intervening absorbers. The reddening was measured from
the shift of the average slope of the quasar spectral energy
distribution between the two samples, which was found to be $\langle
\Delta \alpha \rangle = 0.5$ (significant at above 99\% of confidence
level).

Some recent determinations based on larger samples of quasars
\citep{mur04,ell05} do not confirm the earlier conclusions by
\citet{pei91} and the new limits set on the dust obscuration along the
line of sight are down by one order of magnitude with respect to the
earlier estimates. \citet{mur04} find that $\Delta \alpha \leq 0.2$ at
$3~\sigma$ for a larger sample of absorbers at $\langle z_{abs}
\rangle \approx 3$. This implies a reddening $E(B-V)< 0.02 mag$ (at $3
\sigma)$. Unfortunately, the determination of the dust absorption at
high redshift is intrinsically an indirect measurement and there is
the possibility that the real dust content is higher than these lower
estimates due to observational biases such as sample selection
effects.  Indeed, for a subsample of Damped Lyman $\alpha$ absorbers,
selected by the presence of the $CaII$ absorption line, \citet{wil05}
and \citet{wil06} find a significant evidence of reddening at moderate
redshift ($\langle E(B-V) \rangle \gtrsim 0.1$ at $z_{abs} \approx
1$); similarly \citet{yor06} measure $\langle E(B-V) \rangle$ up to
$0.085$ for $MgII$ absorbers at $z_{abs} \approx 1.4$. Thus, any
modeling of dust effects on high redshift objects will need to take
into account possible selection effects.

Surprisingly, very little consideration is generally given to the
effects of dust obscuration along the line of sight for high-redshift,
non-active galaxies. Clearly, if the effects of dust absorption have
been detected for quasars, it is likely that they will affect every
other object at similar distances. Nonetheless, it is common practice
in observations of high redshift galaxies to consider the obscuration
of dust as a screen localized at the emitter location \citep[e.g.,
see][]{pap01}, adopting a description of the dust properties like the
one used for local starburst galaxies \citep{cal94}.

The primary goal of this paper is therefore to estimate the fraction
of essentially unobscured lines of sight for very high redshift
objects ($z \lesssim 20$) and their average transmission. The
estimation of this quantity is much more robust with respect to
uncertainties and biases in the observed distribution of absorbers
than the measure of the amount of dust in DLA systems (whose
determination is outside the scope of this work). In fact, while a
small number of optically thick absorbers, missed in magnitude limited
surveys, could in principle contain the majority of dust in the
universe, their effect on the average transmission along a random line
of sight would be limited by their covering factor. The CORALS radio
selected survey \citep{ell01} probes $66$ lines of sight with complete
optical follow-up detection. The probability of finding an optically
thick absorber along a random line of sight is therefore below 4.9\%
at 99\% of confidence level (and below 2.9\% at 95\% of confidence
level).

In Section 2 we present our model for the absorbers, that is
calibrated in Section 3 up to redshift $z \approx 5$ on the
measurements from recent Damped Lyman $\alpha$ (hereafter DLA) surveys
{\citep{ell01,pro05}}.  For extrapolation to higher redshift we assume
an unevolving comoving density of Damped Lyman $\alpha$ systems and a
dust to gas ratio decreasing exponentially with redshift (Section
3.1). We also consider a wider set of input parameters to investigate
different parameters extrapolation recipes and to study a higher dust
content that could be missed due to selection effects. In Section 4,
we describe our Monte Carlo code, that allows to characterize the full
probability distribution of absorption which is presented in Section
5. Section 6 sums up.

\section{Absorber modeling}

Given a source at redshift $z_e$, we are interested in modeling the
absorption due to dust residing in DLA systems (i.e. absorption
systems with neutral hydrogen column density, $N_d$, above $2 \cdot
10^{20} cm^{-2}$). Our approach is inspired by the models by
\citet{mol90} and by \citet{mad95} (see also \citealt{fal93}), in
which the absorber distribution is treated as an input parameter that
we calibrate to observations in the next section.

We assume a discrete probability distribution of absorbers
along the line of sight up to redshift $z_e$ with a \emph{separable}
probability distribution in column density ($N_d$) and redshift
($z$). This probability distribution is further assumed to be
Poissonian per unit redshift, so that
\be \label{eq:1}
p(N_d,z)=\phi(N_d) \psi(z),
\ee
with $E[d \psi(z)/dz]=n(z)$ (in this paper the symbol $E_{p}[f]$ means
the expectation value of $f$ under the probability measure $p$, and we
drop the subscript $p$ if it is clear what probability measure
we are referring to). 

We consider the following forms for the column density distribution
$\phi(N_d)$: (i) a gamma function, the best fitting functional form
observationally identified by \citet{pro05}:
\be \label{eq:xiGamma}
\phi(N_d) \propto \left (\frac{N_d}{N_{\gamma}}\right)^{-\alpha_1} \exp{\left(-N_d/N_{\gamma}\right)},
\ee
with $N_d$ in $[N_{min};+\infty]$ and (ii) a power law 
(computationally friendlier in our Monte Carlo approach due to the simple
analytical primitive) in a range $[N_{min},N_{max}]$:
\be \label{eq:xiPower} \phi(N_d) \propto \left(N_d
\right)^{-\alpha_2}.  \ee
The normalization for the function $\phi(N_d)$ is chosen so as to have
$\int dN_d \Phi(N_d) = 1$.

The average number of absorbers per unit redshift $n(z)$ is assumed to
vary as:
\be
n(z) = A dX/dz,
\ee
where 
\be
\frac{dX}{dz}=\frac{H_0}{H(z)}(1+z)^2=\frac{(1+z)^2}{\sqrt{\Omega_{\Lambda}+\Omega_M(1+z)^3}}. 
\ee
Here we adopt a WMAP concordance cosmology with
$\Omega_{\Lambda}=0.7$, $\Omega_{M}=0.3$ and $H_0=70~km/s/Mpc$
\citep{spe06}.

The absorption cross section of dust located at $z_a$ is assumed to
be $\sigma(\lambda_a)$ in its rest frame. For an observed
wavelength $\lambda_o$ this can be written as:
$\sigma(\lambda_o/(1+z_a))$. In the B-band in the absorber rest frame, 
$\lambda_a=0.44 \mu m$, we have that the optical depth of the dust can
be written as:
\be
\tau_B= N_{d}\sigma(0.44 \mu m)= k \frac{N_{d}}{10^{21}cm^{-2}},
\ee
where $k$ is a dimensionless dust to gas ratio parameter ($k$ is of
order unity for the Milky Way).

At different wavelengths the absorption
cross section can be expressed via an extinction curve $\xi(\lambda)$:
\be
\xi(\lambda)=\sigma(\lambda)/\sigma(\lambda_B).
\ee

Thus, for a single cloud at redshift $z_a$ we can write the optical
depth $\tau(\lambda_o)$ as:
\be \label{eq:tauSINGLE}
\tau(\lambda_o) = \frac{k N_d}{10^{21} cm^{-2}} \xi(\lambda_o/(1+z_a)).
\ee

Given these assumptions, we can compute analytically - down to the
numerical evaluation of a single integral over redshift - the average
value for the transmission coefficient $q(\lambda_o,z_e)=\exp{ \{ -
\tau(\lambda_o,z_e)\}}$ to a source at redshift $z_e$. The computation
is straightforward \citep[see Appendix A in][]{mol90} and yields:
%
%\be \label{eq:obs} E[\tau(\lambda_o,z_e)] = \int_0^{z_e} dz ~n(z) k(z)
%E_{\phi}[N_d] \xi(\lambda_o/(1+z)), \ee
%
%\noindent
%and
%
\be \label{eq:Eq}
E[q(\lambda_o,z_e)] = \exp{ \left \{ - \int_0^{z_e} dz ~n(z) (1-
 E_{\phi}[q(\lambda_o,z)]) \right \} }
\ee
with
\be
q(\lambda_o,z) = \exp{ \left \{ - k(z) N_d \xi(\lambda_o/(1+z)) \right \} }.
\ee
We recall that $E_{\phi}[f]$ denotes the expectation value of $f$ under
the probability distribution $\phi$.

From the average value of the transmission coefficient we can define
an \emph{effective} optical depth:
\be \label{eq:taueff} \tau^{(eff)} = -\log{E[q(\lambda_o,z_e)]}.  \ee

This quantity is a measure of the average departure from unit
transmission and is more physically relevant than the mean optical
depth for the purpose of characterizing the fraction of obscured lines
of sight in the sky \citep[see also][]{mad95}. In fact the mean value
of the optical depth is very sensitive to the detailed properties of
the high tail in the $\tau$ distribution. The concept can be easily
illustrated with the following example. Consider 100 lines of sight,
one of which with a really optically thick absorber ($\tau = 10^4$)
while the others are optically transparent. In this case, the
\emph{effective} optical depth (Eq.~\ref{eq:taueff}) is $\tau^{(eff)}
\approx 0.01$ and correctly captures the fact that with probability
$1\%$ a line of sight is optically thick. The average optical depth is
instead $E[\tau]=10$, a very misleading value if applied to the
estimation of the probability of having a line of sight free from
absorption.

\section{Parameter estimation from DLA data}\label{sec:fit}

We can take advantage of recent surveys
\citep{ell01,pro05,ake05,rao06}, that have measured the observed
redshift-column density distribution and metallicity for DLA systems,
to critically examine the assumptions adopted in our model and to
estimate its free parameters.

In principle, the calibration of our model for the absorbers appears
straightforward, as it relies on the observed statistics of Lyman
$\alpha$ features in the spectra of several thousands of quasars over
an extended redshift range. Unfortunately, for optically selected
quasars samples, the observed distribution of DLA systems is in
general a biased estimator of the intrinsic distribution as highly
obscured lines of sight are preferentially missed for an optically
selected sample. However, as we are interested in characterizing the
average transmission to high z, and not in the measure of the comoving
gas density in DLA systems, the dust bias at the high end of the
column density distribution of neutral gas is of limited impact. {In
fact, if a small fraction $\epsilon$ of optically thick absorbers is
missed in a survey, this will only introduce a relative error of order
$\epsilon$ in the average transmission. The comoving gas density may
on the other side be affected by an arbitrarily large error if these
missed absorbers dominate the gas density budget.}

For the calibration we resort to two samples of DLA data. The first is
the radio selected CORALS survey \citep{ell01} that has the advantage
of being free from dust bias, but consists only of 66 quasars with
detection of 19 intervening DLA systems. The statistical uncertainties
in the values of the estimated parameters are rather large (reported
as entries ``El01'' in Table~\ref{tab:model}). We therefore also
consider the significantly larger, but optically selected, SDSS DR3
DLA dataset \citep{pro05} that consists of 525 DLA systems identified
in the spectra of 4568 quasars with signal to noise ratio above
4. This dataset, while providing an excellent statistical accuracy
(the \emph{observed} neutral gas density in DLA systems is measured
with relative error below 10\%; see \citealt{pro05}) may be affected
by systematic uncertainties for the high column density tail of the
DLA systems distribution (see \citealt{tre06} for a characterization
of the systematic errors in the SDSS DR3 DLA dataset). The sets of
parameters estimated using these data are reported as entries ``Pr05''
(SDSS) in Table~\ref{tab:model}.

The absorbers column density distribution $\Phi(N_d)$ has been
characterized starting from the empirical distributions for the CORAL
and SDSS surveys using a maximum likelihood estimator and plotting, in
Fig.~\ref{fig:like}, the likelihood function.

We estimate the values for the parameters in the gamma function
description for $\phi(N_d)$ using only the SDSS data, as the fit would
have too many free parameters for the size of the CORALS survey. With
our analysis we re-derive the same parameters identified by
\citet{pro05} (entry ``Pro05\_$\Gamma$'' in
Table~\ref{tab:model}). Namely, we assume the standard DLA limit
$N_{min}=2 \cdot 10^{20} cm^{-2}$, and we find $\alpha_1 = 1.8$ and
$N_{\gamma} = 3 \cdot10^{22} cm^{-2} $. For the single power law
description we obtain $\alpha_2= 2.2$ (SDSS) and $\alpha_2=2.1$
(CORALS) adopting $N_{min}=2 \cdot 10^{20} cm^{-2}$ (the standard DLA
limit density) and $N_{max} = 10^{22} cm^{-2}$. This cut-off has been
introduced with the goal of eliminating unphysical high $N_d$ tails in
the distribution of gas column densities. The cut-off has been set to
a value marginally higher than all the $N_d$ measurements in the SDSS
and CORALS survey. We also explore different power law models with
increasingly high value for the cut-off (models El01\_a, El01\_d and
El01\_e ) in order to assess the effects of a small additional number
of optically missed absorbers with increasingly high column
densities. While the SDSS data rule out a single power law at a
confidence level above $3 \sigma$ for the observed column density
distribution, this description may still be valid for the intrinsic
distribution. Unfortunately, the CORALS data do not allow to
significantly constrain the functional form of $\phi(N_d)$ (see
\citealt{ell01}). The bias in an optically selected DLA survey maps an
\emph{intrinsic} power law distribution of column densities into an
\emph{observed} gamma function distribution (\citealt{fal93}).
%It should be stressed that for the purpose of
%investigating the effects of dust absorption on the transmission along
%random lines of sight, a precise modeling of the high $N_d$ end in
%$\Phi(N_d)$ is not critically needed, as the fraction of relatively
%unobscured lines of sight is not too sensitive on the details of the
%tail of $\phi(N_d)$ (see Fig.~\ref{fig:pr}).

In our approach, $A$ is assumed to be independent of redshift, i.e. we
are assuming a constant comoving number of absorbers. Due to the
separability of Eq.~\ref{eq:1}, this implies a constant comoving
density of neutral gas in DLA systems ($\Omega_{HI}^{(DLA)}$). The
SDSS data \citep{pro05} show evolution of $\Omega_{HI}^{(DLA)}$ by
about a factor $2$, mainly in the redshift range $[2.2;3]$. However,
at lower redshift, the measured gas density seems to be more in line
with the $z > 3$ values (see Fig.~22 in \citealt{pro05} and Fig.~16 in
\citealt{rao06}), so that this schematic modeling appears in
reasonable agreement with the observations. The combined evolution
seen by \citet{rao06} in the line and column density distributions at
approximately constant $\Omega_{HI}^{(DLA)}$ would affect the
separability assumption of Eq.~\ref{eq:1}. This would complicate the
numerical treatment of our model, but would not significantly change
the value of the effective extinction to a given redshift, that, under
the assumption of optically thin absorbers, depends in first
approximation only on the comoving dust (i.e. neutral gas)
density. The average transmission would depend more critically on the
precise form of Eq.~\ref{eq:1} if a significant population of
optically thick absorbers is present, but this is an unlikely scenario
given the complete optical follow-up detection in the radio selected
CORALS survey \citet{ell01}.

In order to estimate $A$, we have tested the covering factor of DLA
systems.  Starting from the published data we have identified for each
quasar in the two samples the range of redshift in which the presence
of $DLA$ systems was detectable (from Table~1 in \citealt{pro05} and
from Table~3 in \citealt{ell01}), computed the expected number of DLA
systems in that interval and obtained $A$ by evaluating the likelihood
of getting the observed number distribution of DLA counts.

The likelihood for $A$ given the two datasets is reported in
Fig~\ref{fig:like}: the maximum value for CORALS is at $A=0.0910$,
while the maximum for SDSS is at $A=0.0715$. Assuming the SLOAN value
for $A$, the CORALS data are marginally consistent: the higher CORALS
result may be due to small number fluctuations at the $1\sigma$ level
(see the likelihood ratio in Fig.~\ref{fig:like}). An alternative
possibility is that the discrepancy is due to an obscuration bias for
SDSS. This is however highly unlikely, as the covering factor
determination is dominated by DLA systems at the low end of the column
density distribution, where the obscuration bias is negligible. The
$A$ value estimated from the SDSS data could even be an overestimate
for the intrinsic $A$ because of a Malmquist bias. That is, more
absorbers with column density below the DLA limit have been scattered
into the DLA sample than absorbers above the limit have been scattered
out \citep[][Prochaska: private communication]{ome06}.

For these parameters, we can compute the comoving density of neutral
gas in DLA systems implied by our model (e.g. see \citealt{pro05}):
\be
\Omega^{(DLA)}_{HI}=  \frac{\mu m_H H_0}{c \rho_c} A \cdot E_{\Phi}[N_{d}],
\ee
where $m_H$ is the mass of the hydrogen atom, $\mu=1.3$ is a
correction factor for the composition of the gas, $c$ the speed of
light and $\rho_c$ the critical density of the universe. We have:
$\Omega^{(DLA)}_{HI} = 0.81 \cdot 10^{-3}$ for SDSS fitted to a gamma
function, $\Omega^{(DLA)}_{HI} = 0.84 \cdot 10^{-3}$ for the SDSS
fitted to a power law and $\Omega^{(DLA)}_{HI} = 1.17 \cdot 10^{-3}$
for CORALS (with $A=0.0910$). The agreement with the
published value from SDSS is excellent (see Table~9 in
\citealt{pro05}: their unbinned measurement is $(0.817 \pm0.05) \cdot
10^{-3}$), while our data cannot be compared directly with the
analysis in \citet{ell01}, as they have used a different cosmology
($\Omega_M=1$ and $\Omega_{\Lambda}=0$).

The value of the dust to gas ratio $k$ for DLA systems depends on
their metallicity $Z$ and as a first approximation we can consider a
linear dependence of the dust to gas ratio on $Z$. DLA systems are
generally characterized by a low metallicity and by a moderate
evolution of their properties with the redshift \citep{wol05}. Their
metallicity has been measured in several surveys and the average
metallicity in optically and radio selected samples appears consistent
\citep{ake05}. Here we consider the compilation by \citet{kul05} {(see
also \citealt{pro03})} and we approximate the reported measurements
(Fig.~13 in \citealt{kul05}) with a linear function for
$\log{(Z(z))}$, which provides a good agreement with the data in the
redshift range $2 \lesssim z \lesssim 5$:
\be \label{eq:zz} Z(z)/Z_{\sun}=0.2 \cdot 10^{-0.2z}. \ee
By considering a typical dust to gas ratio $k=0.8$ for our galaxy ($Z
\approx Z_{\sun}$, with $\approx 50\%$ of the metals locked in dust
grains), this translates into an observed dust to gas ratio:
\be \label{eq:ZKul} k(z)= 0.16 \cdot 10^{-0.2z}. \ee
To account for uncertainties in the measure of $Z(z)$ and for a
smaller fraction of metals depleted into dust, we introduce a free
factor $\alpha_{\kappa}$ for the intrinsic dust to gas ratio $k(z)$:
\be
k(z)=\alpha_{\kappa} \cdot k_o(z).
\ee
Realistic values for $\alpha_{\kappa}$ range in the interval
$[0;1]$. $\alpha_{\kappa}=0.5$ corresponds to $25\%$ of the total
metal amount in dust grains (e.g., see \citealt{pet97},
\citealt{pro02}), while $\alpha_{\kappa}=1$ implies a depletion factor
like in the Milky Way \citep{pei95,pei99}. Our main results are
presented in terms of $\alpha_{\kappa}=1$, so that we effectively
obtain upper limits on the obscuration along the line of sight. In
the following Sections we will also discuss scenarios with
$\alpha_{\kappa}<1$.

The extinction curve $\xi(\lambda)$, in the absorber rest-frame, is
assumed to be as measured in and parameterized for the Small
Magellanic Cloud \citep{pei92}. In fact, {extinction curves similar to
those measured for our Galaxy and for the Large Magellanic Cloud} seem
to be ruled out by the current observational data \citep{ell05,yor06};
extinction curves for DLA systems at low redshift start to be directly
measured and highlight a rather complex picture: \citet{jun04} have
measured $\xi(\lambda)$ for a $z=0.524$ DLA absorber finding some
similarities with the Galactic extinction curve. For our purposes the
use of a different extinction curve would not affect our results
significantly, as the redshift averaging process over many absorbers
tends to smooth out the specific features of the input curve. 

\subsection{Extrapolation of parameters up to $z \approx 20$}\label{sec:DLAsam}

As we are mainly interested in characterizing the expected absorption
for future observations at $z>6$, the parameters that have been tuned
to the properties of DLA systems at $z \lesssim 5$ have to be
extrapolated into a redshift region with no observational constraints.

Qualitatively, a monotonic metal (and dust) abundance appears plausible
even before reionization when all hydrogen is neutral. Indeed, as the
redshift increases the metallicity decreases, so one expects that the
average local content of dust will progressively be reduced.

Our derivation of the average transmission depends on the comoving
dust distribution which we treat as the product of dust-to-gas ratio
times neutral hydrogen distribution in discrete systems which we
identify as DLA systems at $z<5$ but that could simply be the sites of
metal production at higher z.

At a sufficiently large $z$ the average comoving density of neutral
gas will stay approximately constant: the star formation rate should
drop after $z \approx 6$ and this, combined with progressively less
time available for star formation, means that the initial reserve of
gas should remain almost undepleted. Eventually, in a hierarchical
formation scenario, one can expect that the neutral gas will reside in
more numerous Lyman $\alpha$ systems with smaller column densities,
but this will influence only marginally the value of the expected
absorption along the line of sight {at constant comoving gas
density. In fact, given their very low dust-to-gas ratio, these
primordial absorbers are expected to be optically thin, so that the
average transmission will not depend in first approximation on the
details of the distribution and will be proportional to the integrated
comoving dust density.}

These arguments led us to choose to extrapolate our set of parameters
with minimal assumptions: constant $A$, i.e. constant comoving
$\Omega_{HI}$ density, and exponentially decreasing metallicity, as
given by extension of our Eq.~\ref{eq:ZKul}. These extrapolations are
consistent with the asymptotic behavior of global models for the
chemical evolution of DLA systems, like the ones developed
by \citet{pei95}.

The use of Eq.~\ref{eq:zz} implies a metallicity of $1.2 \cdot 10^{-3}
Z_{\sun}$ at $z=11$ which is the redshift of reionization derived from
the WMAP 3 year Compton optical depth \citep{spe06}. This value is
comfortably larger than the minimum metallicity required for
reionization (by PopIII stars, $Z \approx 1.2 \cdot 10^{-4} Z_{\sun}$,
\citealt{sti04}). This extrapolation could thus be considered as
conservative.

An additional test can be performed in terms of the predictions given
by models for the formation and chemical enrichment of DLA systems
based on cosmological hydro-dynamical simulations \citep{cen03} or on
semi-analytical prescriptions \citep{joh06}. These models confirm a
relatively slow but progressive decrease of the metallicity, while
they evidence a sharp drop in the comoving density of neutral gas in
DLA systems at $z>7$ \citep[][private communication]{joh06}, that
however could be due to the built-in assumptions, e.g. coeval
evolution with equal ages.

To check what would be the consequences for our estimates in a
scenario where the number of absorbers drops significantly at $z>7$,
we have run some Monte Carlo simulations assuming that there are no
absorbers at $z>7$ and compared the obscuration given by these models
with that predicted by those with extrapolation at constant $A$. The
results - presented in detail in Sec.~\ref{sec:eff_abs} - evidence a
modest variation in the average transmission. This is easily
understood as, given the exponential decrease of $k$, the impact of
$z>7$ absorbers is limited.

\section{Our Monte Carlo Code}

In order to compute the probability distribution for the
optical depth we resort to a
Monte Carlo code. 
Our code accepts general input functions $\Phi(N_{d})$, $n(z)$, $k(z)$
and $\xi(\lambda)$ and generates the chosen number of discrete
realization for the DLA system distribution along lines of sight for a
given emission redshift $z_{e}$. The cumulative probability
distribution for the optical depth to redshift $z_e$ is then built.

For each discrete realization, we begin by integrating up to $z_e$ the
redshift distribution of DLA systems $n(z)$ so as to obtain the expected total number of absorbers along the
line of sight:
\be n_{tot}=\int_{0}^{z_e} n(z) dz.  \ee
A Poisson random variable with mean $n_{tot}$ representing the
realized number of absorbers is generated using a standard subroutine
from \citet{NR}. The redshift for each absorber is then randomly
assigned by inversion of the primitive of $n(z)$, evaluated
numerically.  Similarly the column density for each absorber is
assigned via random sampling by inversion of the primitive for
$\Phi(N_d)$. Once the redshift and column density for each absorber
has been assigned, the optical depth at the observed wavelength of
interest is computed for each absorber via Eq.~\ref{eq:tauSINGLE}. The
total optical depth is given by summing over all the absorbers along
the line of sight.

The accuracy of the code is estimated by evaluating the variance for
selected levels in the optical depth distribution (median, upper and
lower $1,2 \sigma$ points). In addition the effective optical depth is
compared to the analytical value from Eq.~(\ref{eq:Eq}). The effective
optical depth and the median are evaluated, at a fixed number of
discrete realizations, with greater accuracy than the $1$ and $2
\sigma$ points. Especially for the $2 \sigma$ contours, one order of
magnitude more realizations are needed for an accuracy comparable to
the one reached for the effective optical depth. For our purposes we
are satisfied with an absolute error below $10^{-3}$ for
$\tau^{(eff)}$. This is reached with about $10^5$ random lines of
sight. Depending on the redshift of emission (lines of sight for low
$z_e$ have an expected number of intervening DLA system much less than
one, so the relative variance in the MC code is higher) this
translates into a relative error below $10^{-2}$ at low $z_e$ and of
about $2-3 \cdot 10^{-3}$ at high $z_e$ for the MC simulations
presented in this paper (see Fig.~\ref{fig:MC}).

\section{Results: absorption probability distribution}\label{sec:eff_abs}

In Figs.~\ref{fig:El01_a}-\ref{fig:El01_b} we plot the effective
optical depth $\tau^{(eff)}$ (at the emitter rest-frame
$\lambda_e=0.14~\mu m$) to a given redshift, obtained using our models
El01\_a and El01\_b, which employs the best fitting value for $A$ from
the CORALS survey (that represents a generous upper limit for $A$, see
Sec.~\ref{sec:fit}) and, respectively, $\alpha_{\kappa}=1$ (50\% of
metals in dust) for model ``a'' and $\alpha_{\kappa}=0.5$ (25\% of
metals in dust) for model ``b''.

The effective optical depth peaks at about $z \approx 5$ with a
maximum value below $\tau^{(eff)} \lesssim 0.08$ for the model with
the highest dust-to-gas ratio. As the emission redshift increases the
optical depth at fixed emitted wavelength decreases. In fact, despite
the increase in the redshift density $n(z)$, high $z$ absorbers are
characterized by a lower metallicity, decreasing exponentially with
$z$ in our model, while absorbers at lower $z$ are traversed by light
that has been redshifted toward progressively longer wavelengths,
where the absorbers are more transparent. This explains the shape of
the differential contribution to the effective optical depth shown in
Fig.~\ref{fig:integ}. Even for $z>10$ observations the main
contribution to absorption comes from systems at $2 \lesssim z
\lesssim 5$, which is a range probed with the highest accuracy by
current DLA surveys.

The optical depth distribution is characterized by a small number of
highly obscured lines of sight, while the vast majority is almost
dust-free: e.g., for our standard model (El01\_a), along a random
direction $\tau<0.1$ with probability $\approx 0.8$. Only $5\%$ of the
lines of sight may have $\tau \gtrsim 0.35$ to $z=5$, while to $z=20$
we have $\tau<0.1$ with probability $0.95$. In the El01\_b model the
optical depth is below $0.2$ with probability 0.95 at $z \approx 5$
and declines below $0.1$ for $z \gtrsim 12$ with probability
0.95. {The optical depth distribution
(Figs.~\ref{fig:El01_a}-\ref{fig:El01_b}) shows a sharp rise from 0 to
the value of the minimum optical depth for a single absorber at the
redshift where the probability of having a clear line of sight falls
below the probability associated to the line plotted.}

Decreasing the dust-to-gas ratio (models El01\_a-El01\_c) leads to a
corresponding quasi-linear decrease in the effective optical depth
(see Eq. \ref{eq:Eq} and {compare Fig.~\ref{fig:El01_b} - model
El01\_b - with Fig.~\ref{fig:El01_a} - model El01\_a }).

Even if we consider alternative models the expected effective optical
depth does not change dramatically. In Fig.~\ref{fig:pr} we report the
optical depth distribution for our models calibrated to the SDSS DLA
survey data. The two different models considered (Pr05\_P -power law-
and Pr05\_$\Gamma$ -gamma function- for $\Phi(N_d)$), have negligible
differences between each other in terms of the resulting effective
optical depth. The SDSS data suggest an effective optical depth that
is about 30\% smaller than that from the CORALS data.

In Fig.~\ref{fig:comparison} we explore the effects of variations of
different assumptions for the modeling of the absorber
distribution. One important parameter that is difficult to constraint
observationally is the cut-off value $N_{max}$ for the power law form
of $\Phi(N_d)$ used to fit the CORALS data.  We have considered two
additional models (El01\_d and El01\_e) with cutoffs two and ten times
higher than El01\_a (see Tab.~\ref{tab:model}). The effective optical
depth, shown in Fig.~\ref{fig:comparison}, changes by $\Delta
\tau^{(eff)} \lesssim 0.02$ going from El01\_a to El01\_e. Absorbers
with high dust column densities, that may have been missed in the
CORALS survey due to small number statistics (and that are likely to
be missed in optically selected surveys like SDSS), have only a modest
effect on the expected average transmission. {One caveat is that this
conclusion has been reached by extrapolating a power law column
density distribution into a region of the parameter space ($N_{d}>
10^{22} cm^{-2}$) where there are no observational constraints. These
systems may well be a distinct population of absorbers with column
density and dust-to-gas ratio distributions different from those of
the observed DLAs. In particular an absorber with $N_{d}> 10^{22}
cm^{-2}$ may contain $H_2$ and hence have an higher dust-to-gas
ratio. An upper limit to the uncertainty introduced on the expected
average transmission by a hypothetic population of ``bricks''
absorbers can be estimated from the CORALS survey. At $95\%$ of
confidence level this population influences less than $2.9 \%$ of the
lines of sight. At this confidence level the maximum displacement
introduced in $\tau^{(eff)}$ is $\Delta \tau^{(eff)} \lesssim 0.03$,
in good agreement with the estimate $\Delta \tau^{(eff)} \lesssim
0.02$ that we obtain with our El01\_e model (for the effects of
dust-rich absorbers see also the model El01\_h discussed below).}

Model El01\_f, shown in Fig.~\ref{fig:comparison}, is
characterized by a cut-off of the absorbers distribution at $z=7$,
while otherwise coincides with the model El01\_a. This allows to
quantify the uncertainties associated with the extrapolation of our
models to redshifts where DLA data are unavailable. The possibility
that the DLA number density may drop significantly at high redshift is
hinted by semi-analytical models \citep{joh06}. Even in the extreme
scenario of model El01\_f, where the DLA number is set to 0 for $z>7$,
the difference in the average effective optical depth is only $\Delta
\tau^{(eff)} \lesssim 0.008$.

Model El01\_g investigates the effects of the presence of a
sub-population of DLA systems at $z \lesssim 1.8$ with high
metallicity \citep{wil05,wil06}. This model has been constructed
starting from the standard El01\_a and assuming that one third of the
DLA systems at $z<1.8$ have solar metallicity. The presence of a
significant number of these systems at higher $z$ is unlikely, as the
metallicity in the CORALS survey is significantly sub-solar and
consistent with that measured in optically selected surveys
\citep{ake05}. The effect of this population of absorbers is to
enhance the optical depth up to $z \lesssim 6$ (see
Fig.~\ref{fig:comparison}). However as the emission redshift further
increases their influence is significantly reduced and becomes
negligible for $z \gtrsim 10$. 

Model El01\_h continues to investigate the effects of a small
population of optically thick absorbers (like the inner regions of
spiral galaxies or dusty star bust galaxies like M82) that could have
been missed in the CORALS survey due to the limited number of lines of
sight probed. We assume that a random line of sight intersects a
number of galaxies drawn from a Poisson distribution with average
$0.025$\footnote{
This value for the covering factor has been estimated as sum of two
contributions at high and low redshift. For the high redshift
contribution we have analyzed the Hubble Ultra Deep Field. In the
$i_{775}$ band the covering factor of pixels brighter than $m_i=33$ is
$\approx 0.01$. This number is broadly compatible with what is derived
from the luminosity function of Lyman Break galaxies in the redshift
range $2 \leq z \leq 6$ \citep{ste99,fer04,bou05} and extrapolated
down to $z=1$. To estimate the covering factor due to galaxies at
$z<1$ we have considered four SDSS random fields in the $i$ band for a
total area of $\approx 0.15~deg^2$. For each field we have applied a
cut at $+2.5 \sigma$ from the sky level and then removed isolated
pixels. This analysis provides an estimate of the covering factor of
$\approx 0.015$, with single field values in the range
$[0.008;0.022]$.}.
We assume that each galaxy introduces an optical depth of $0.5$
(estimated from \citealt{hol05}). The results of the Monte Carlo
simulation, shown in Fig.~\ref{fig:el01k}, indicate that the effective
optical depth is slightly higher in this case (i.e. the average
transmission is marginally lower with respect to El01\_a). At the
level of the optical depth distribution, only the contour lines
associated to transmission $\ll 1$ are influenced, i.e. those directly
affected by lines of sight intersecting a galaxy. Differences between
the El01\_h and the El01\_a models appear significant only for the top
10~\% of the distribution.

\subsection{Reddening}

The shape of the effective extinction curve $E[\xi(\lambda)] \equiv
\log E[q(\lambda,z_e)]/ \log E[q(\lambda_B,z_e)]$ is only marginally
dependent on the model used or on the emission redshift considered
(see Fig.~\ref{fig:xi}). We can empirically fit in the range
$\lambda_e \in [0.1 \mu m, 8 \mu m]$ the effective average extinction
curve for emitters at redshift $z \gtrsim 1$ with a simple power law
in the form:
\be \label{eq:xi} E[\xi(\lambda_e)] = \eta - \left
(\frac{\kappa}{\lambda_e}\right)^{\zeta}, \ee
with $\eta= -0.55 $, $\kappa = 0.755~\mu m$ and $\zeta = 0.87$.

With this ``universal'' extinction curve, we can infer the typical
values for the average reddening, e.g. $E(B-V)$, that can be estimated
as a fixed fraction of the effective optical depth $\tau^{(eff)}$ at a
reference wavelength.

\section{Discussion}

We present a model for quantifying the effects of absorption due to
dust in $DLA$ systems along the line of sight for sources up to
$z=20$.

The effective optical depth to a given redshift as a function of the
emitted frequency $\lambda_e$ can be evaluated analytically by using
Eq.~(\ref{eq:Eq}). This allows to obtain immediately an order of
magnitude estimate of the effects of the dust absorption on the
average transmission for the class of observations one is interested
in.
%The average optical depth is a
%robust quantity with respect to changes in the functional form of the
%probability distribution $p(N_d,z)$ (Eq.~\ref{eq:1}) at fixed comoving
%dust density. 
For a better characterization of the effects of the extinction, we
study by means of a Monte Carlo method the distribution of the optical
depths to a given redshift, setting upper and lower limits on the dust
extinction.
% that are more sensitive to the specific
%form of the probability distribution $p(N_d,z)$.

Under the reference scenario, that accounts for a large fraction of
metals in dust grains (model El01\_a with 50\% of the metals in dust)
the effects of dust obscuration remain modest even for very high
redshift, with an optical depth at $\lambda_e=0.14 \mu m$ below $0.4$
with probability $0.95$ for $z_e \approx 3$. As the emission redshift
increases the optical depth decreases and for $z_e \gtrsim 15$ our
modeling predict $\tau \lesssim 0.1$ with probability 0.95. We have
explored several alternative possibilities for the input parameters
finding that the effective optical depth varies within a factor 2 at
most even when a population of optically thick absorbers like the
central parts of star-forming galaxies is taken into account.
Therefore the loss of sensitivity and the effects of reddening are not
expected to significantly influence high-z observations with the James
Webb Space Telescope.

In the future we plan to extend the present framework to include
additional effects on the transmission along the line of sight, such
as gravitational lensing magnification and de-magnification, that may
be significant for explaining the observed number counts of bright
quasars in the SDSS Damped Lyman $\alpha$ survey \citep{pro05}.

\acknowledgements

{It is our pleasure to thank Mike Fall, Harry Ferguson, Benne Holwerda
and Jason Xavier Prochaska for interesting discussions and valuable
comments. We also thank Peter Johansson for providing additional
unpublished data from his semi-analytical model for formation and
evolution of DLA systems. We are grateful to the referee for
constructive suggestions that have improved the paper. This work was
supported in part by NASA JWST IDS grant NAG5-12458.

%%%%%%%%%%%%%%%%%%%%%%%%%%%%%%%%% 

\clearpage

%%%%%%%%%%%%%%%%%%%%%%%%%%%%%%%%%%%%%%%%%%%%%
\begin{figure}
  \plottwo{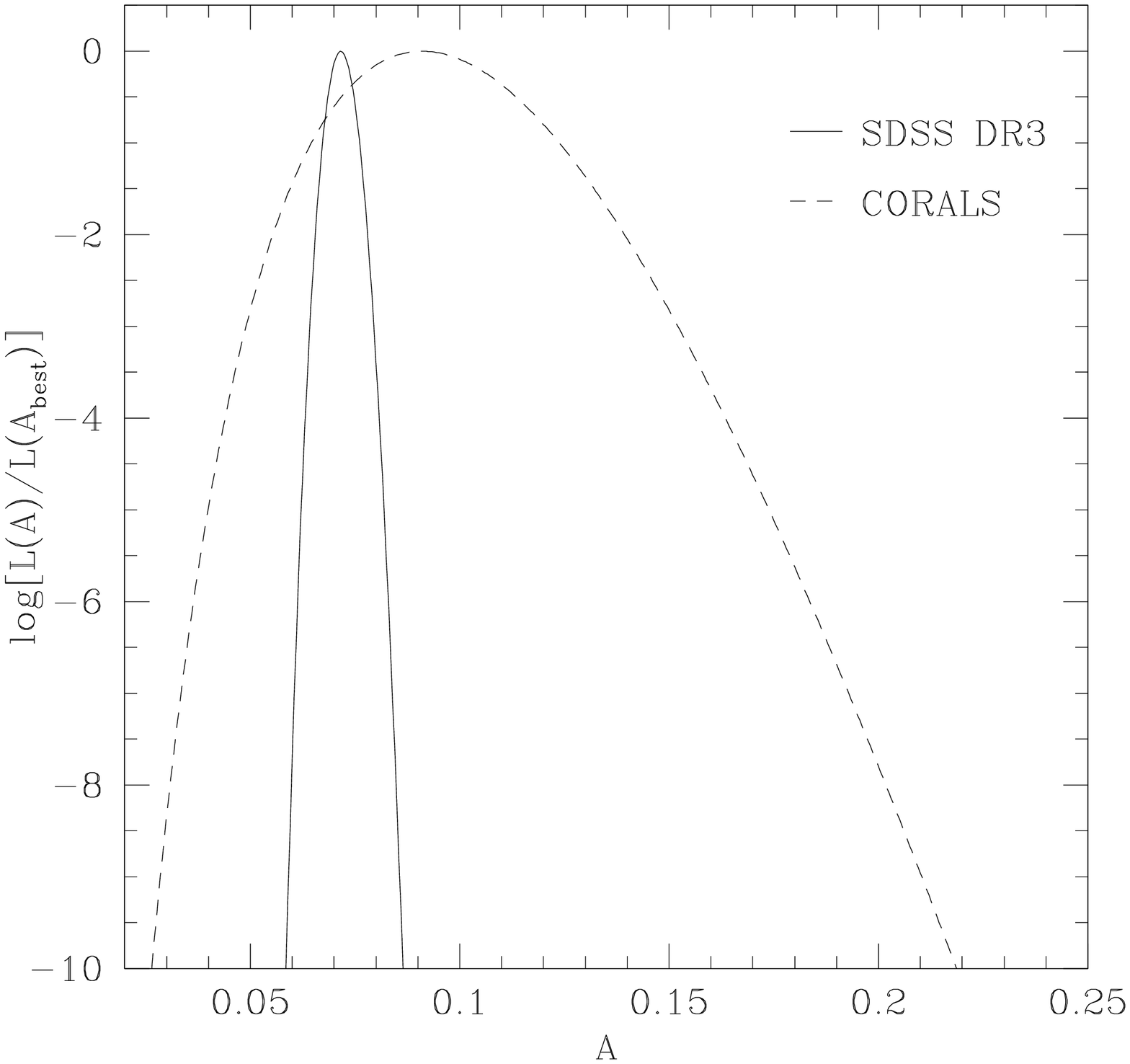}{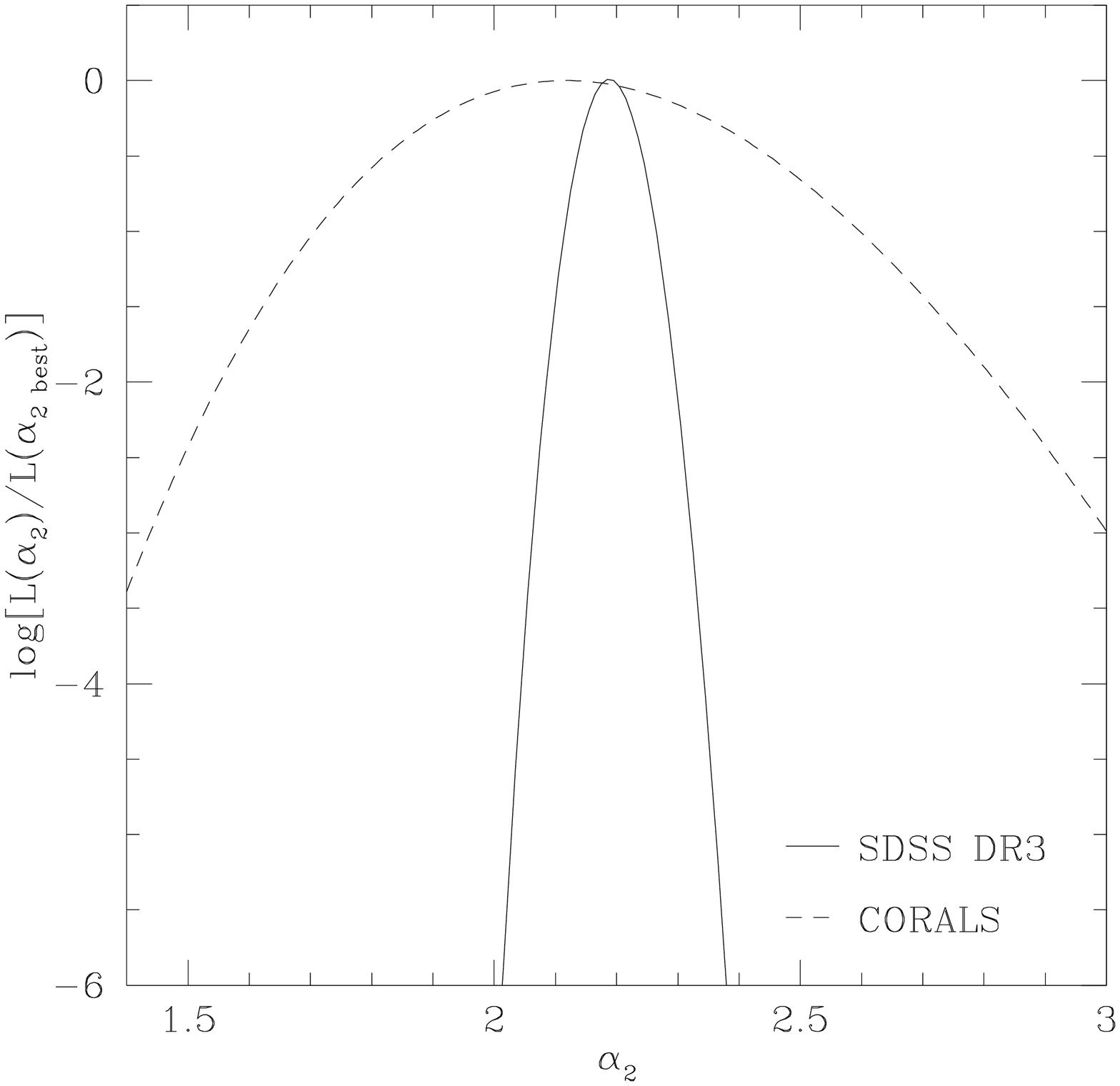} \caption{Maximum likelihood
  estimation for the parameters $A$ (left panel) and $\alpha_2$ (right
  panel), based on the data from \citet{ell01}, dotted line, and from
  \citet{pro05}, solid line. For each dataset we plot the likelihood 
  $L$ (log scale) of the parameters normalized to the maximum
  value.}\label{fig:like}
\end{figure}
%%%%%%%%%%%%%%%%%%%%%%%%%%%%%%%%%%%%%%%%%%%%%%%%

\clearpage

%%%%%%%%%%%%%%%%%%%%%%%%%%%%%%%%%%%%%%%%%%%%%
\begin{figure}
  \plottwo{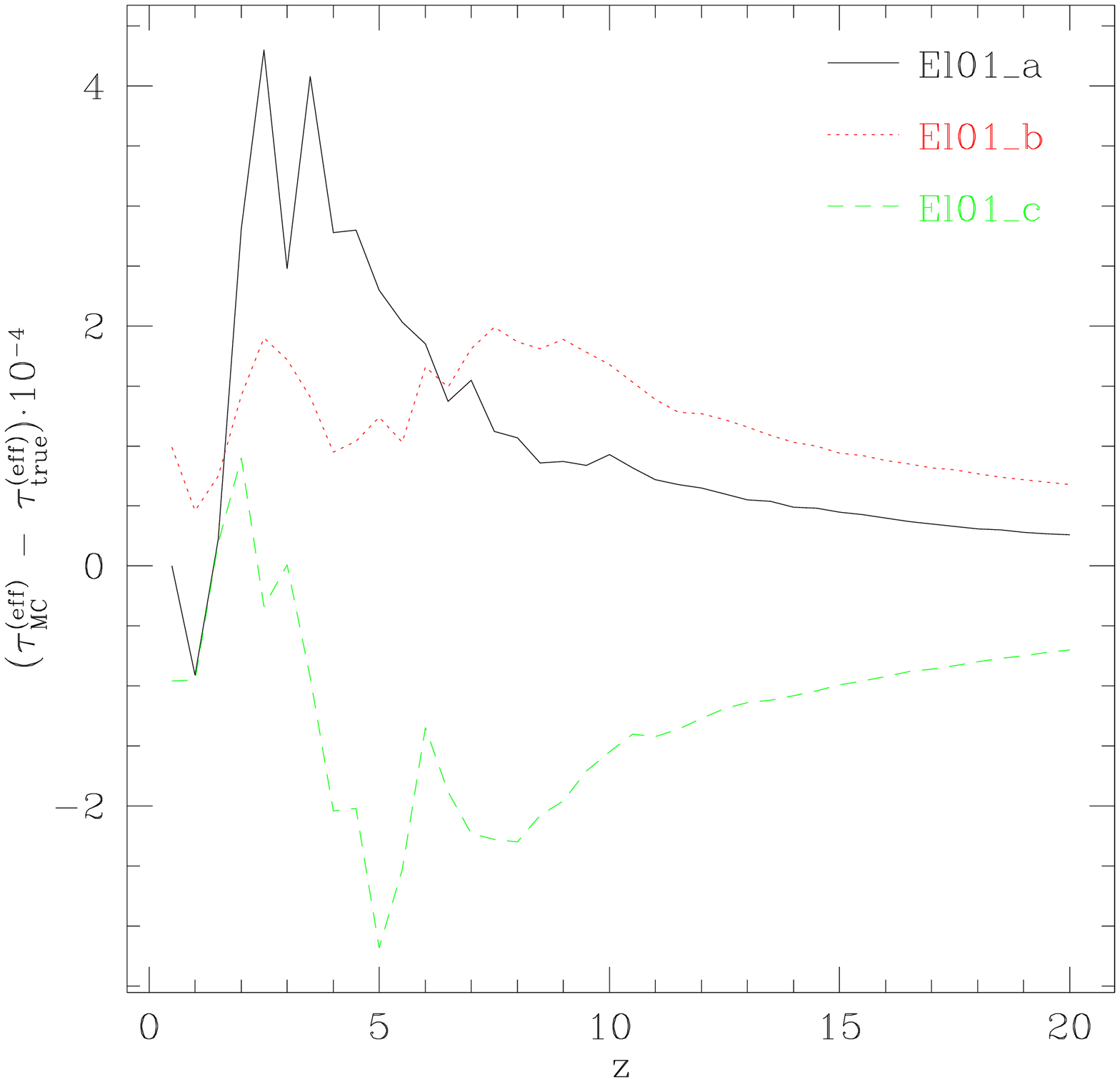}{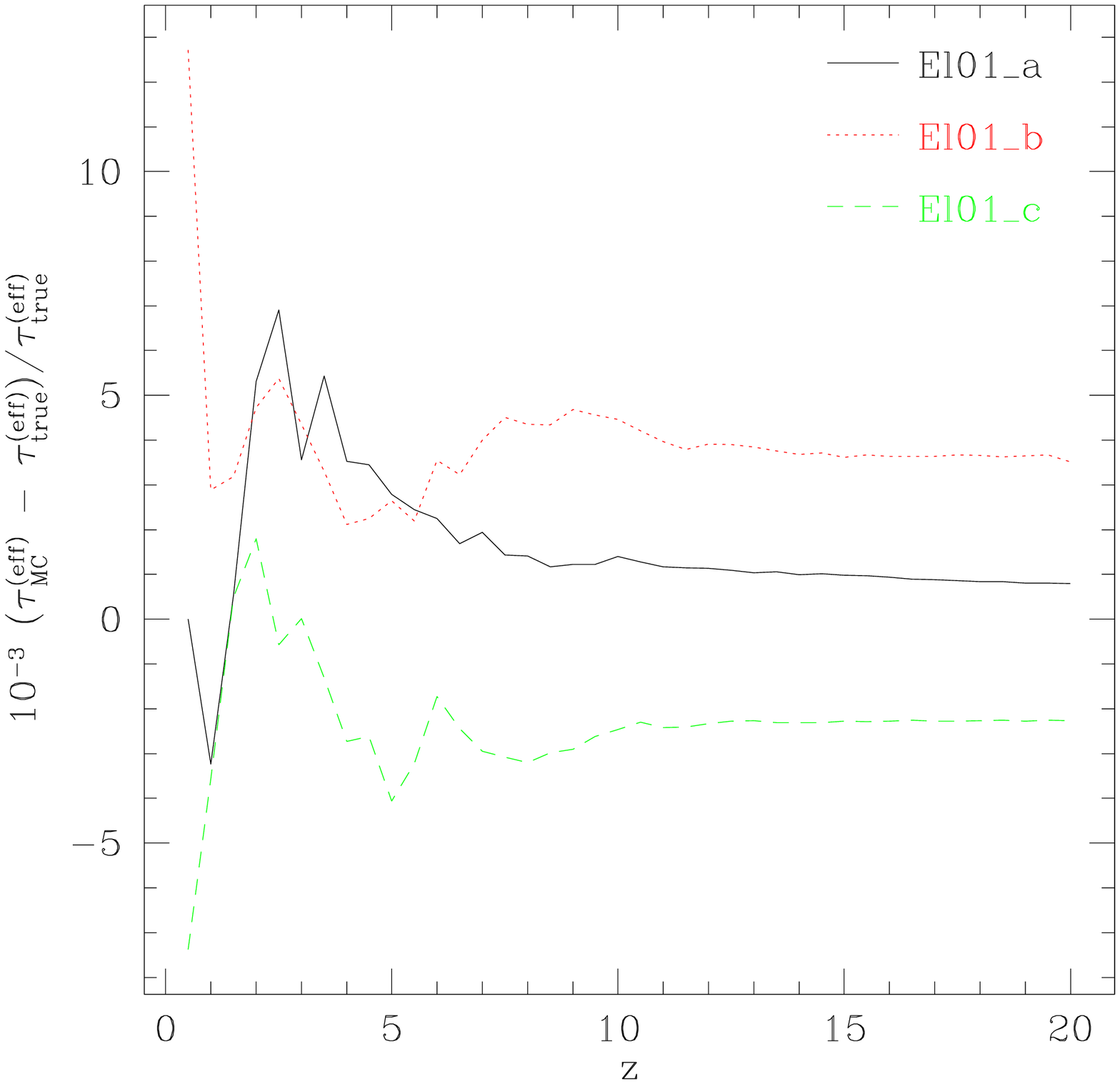} \caption{Absolute (left) and relative
  (right) error for the effective optical depth computed via our MC
  method using $4~\cdot 10^{5}$ random lines of sight and compared
  with the analytical value from Eq.~\ref{eq:Eq} for the models
  El01\_a, El01\_b and El01\_c.}\label{fig:MC}
\end{figure}
%%%%%%%%%%%%%%%%%%%%%%%%%%%%%%%%%%%%%%%%%%%%%%%%

\clearpage

%%%%%%%%%%%%%%%%%%%%%%%%%%%%%%%%%%%%%%%%%%%%%
\begin{figure}
  \plottwo{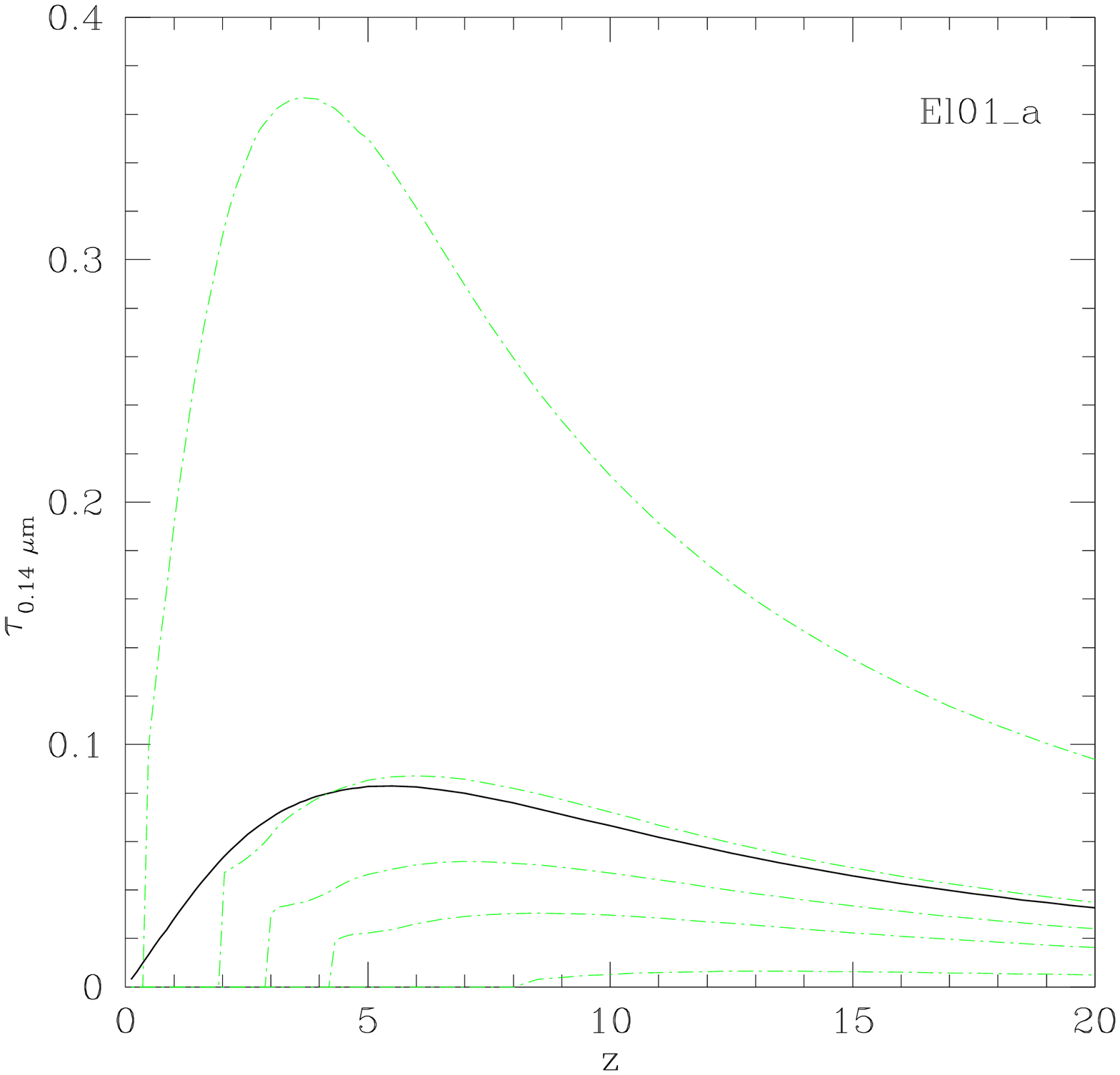}{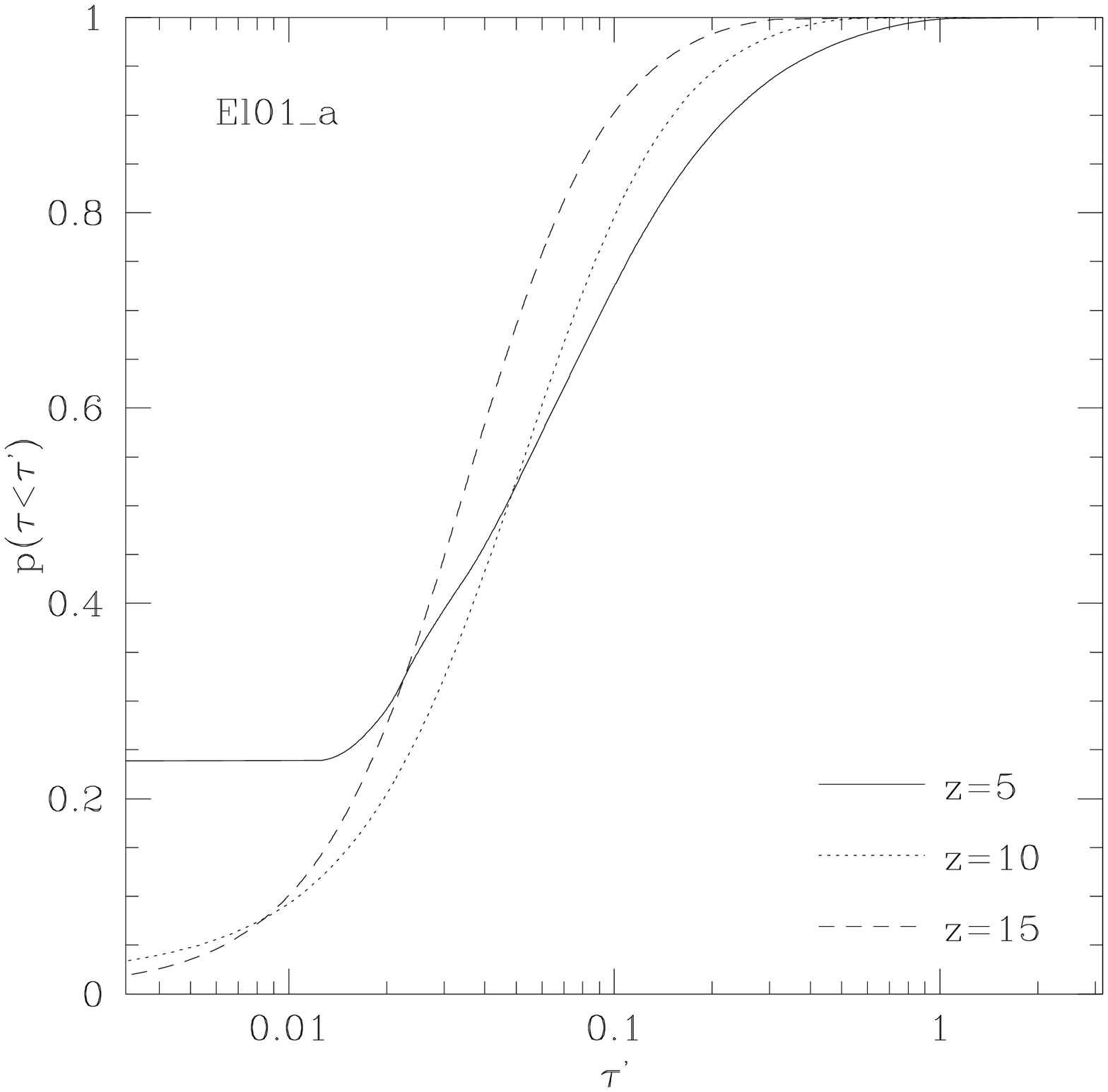} \caption{Left Panel: Effective
  optical depth $\tau^{(eff)}$ as a function of redshift at
  $\lambda_e=0.14 \mu m$ (bold line, solid) for the El01\_a model.
  The green dashed-dotted lines represent, starting from the top in
  the left panel, the upper $95\%$ contour in the distribution of
  optical depth, the upper $68\%$, the median, the lower $32\%$ and
  the lower $5\%$. Right Panel: Cumulative probability distribution
  for the optical depth at $\lambda_e=0.14 \mu m$ along lines of sight
  to different redshifts (solid $z=5$, dotted $z=10$, dashed
  $z=15$). The curves have been generated with a MC code using $4
  \cdot 10^5$ random lines of sight.}\label{fig:El01_a}
\end{figure}
%%%%%%%%%%%%%%%%%%%%%%%%%%%%%%%%%%%%%%%%%%%%%%%%

\clearpage

%%%%%%%%%%%%%%%%%%%%%%%%%%%%%%%%%%%%%%%%%%%%%
\begin{figure}
  \plottwo{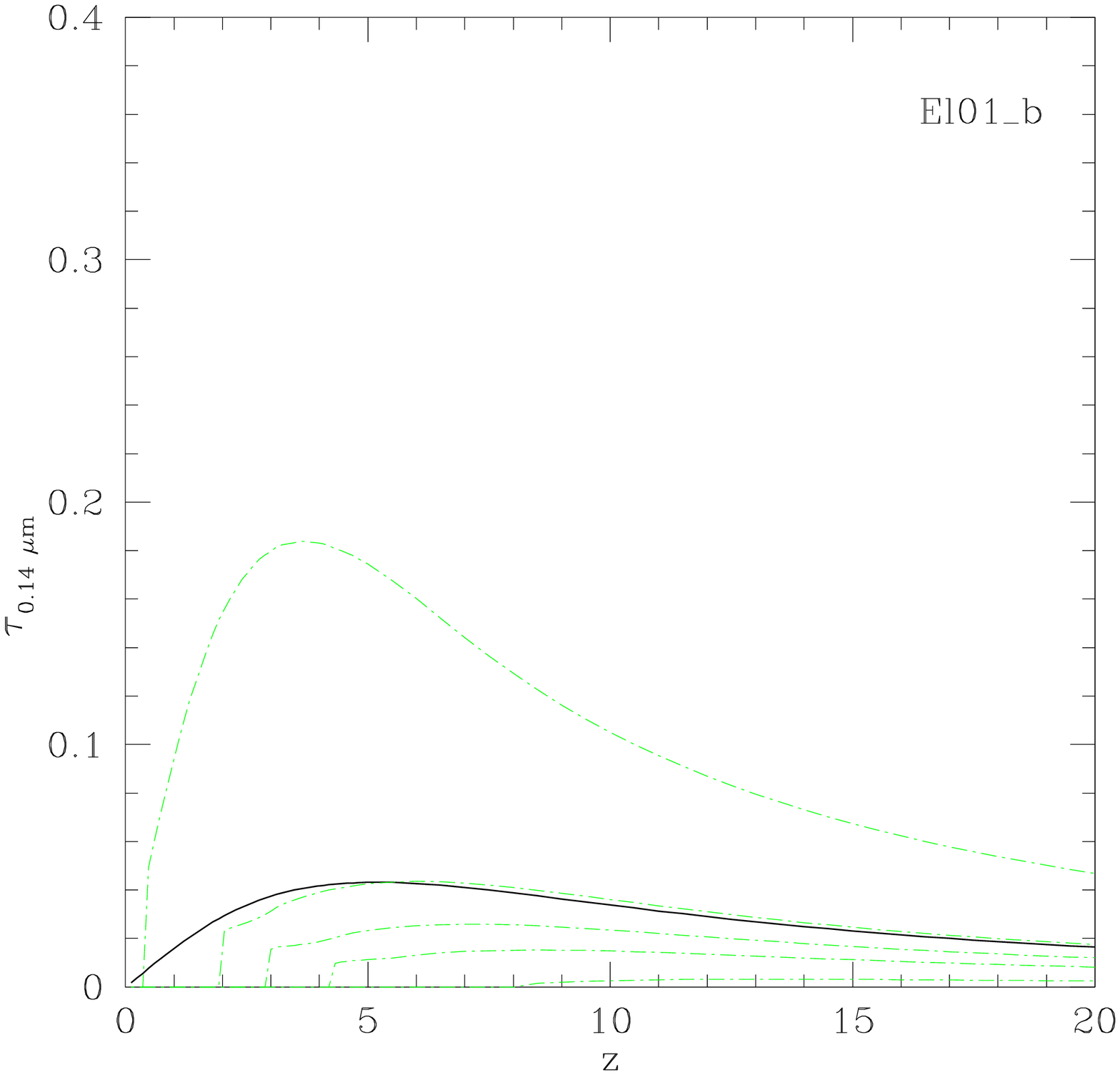}{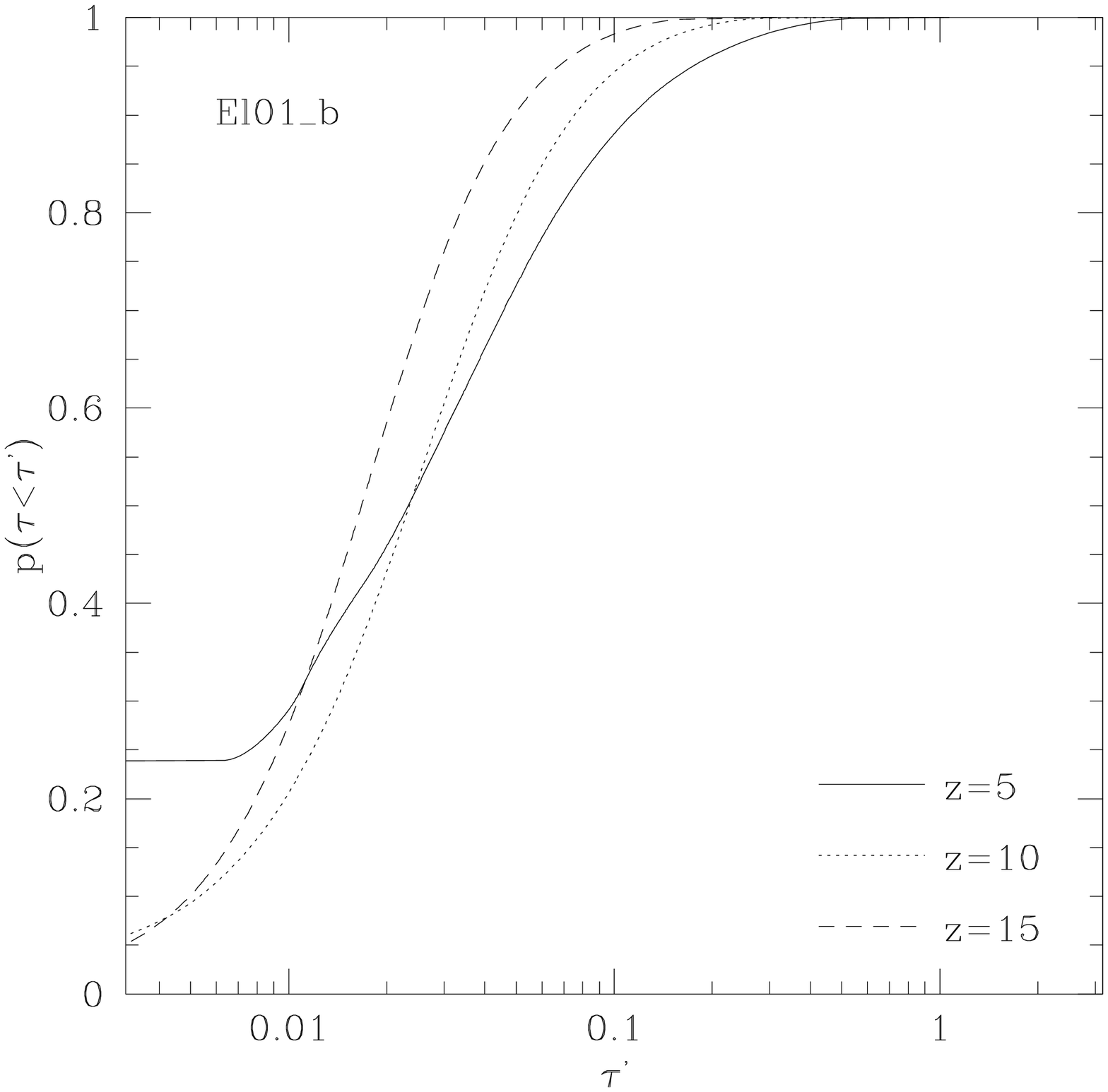} \caption{Effective optical depth
  (left) and cumulative probability distribution for the optical depth
  (right) as in Fig.~\ref{fig:El01_a}, for the El01\_b
  model.}\label{fig:El01_b}
\end{figure}
%%%%%%%%%%%%%%%%%%%%%%%%%%%%%%%%%%%%%%%%%%%%%%%%

\clearpage

%%%%%%%%%%%%%%%%%%%%%%%%%%%%%%%%%%%%%%%%%%%%%
\begin{figure}
  \plotone{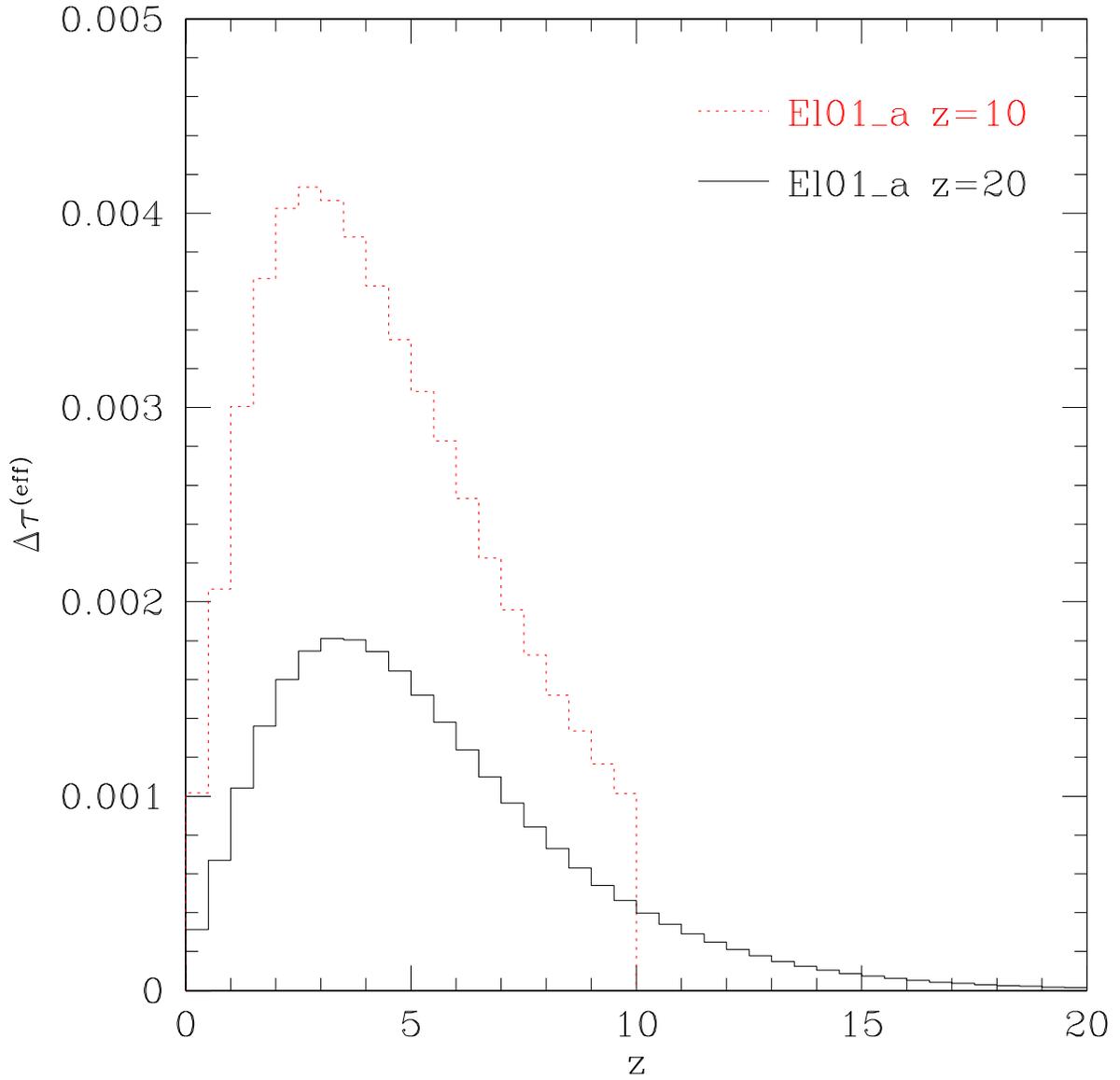}\caption{Contribution to the integral in
  Eq.~(\ref{eq:Eq}) divided in redshift bins. The main contribution
  to the effective optical depth to high redshift sources is given by
  absorbers at $2 \lesssim z\lesssim 5$.}\label{fig:integ}
\end{figure}
%%%%%%%%%%%%%%%%%%%%%%%%%%%%%%%%%%%%%%%%%%%%%%%%

\clearpage

%%%%%%%%%%%%%%%%%%%%%%%%%%%%%%%%%%%%%%%%%%%%%
\begin{figure}
  \plotone{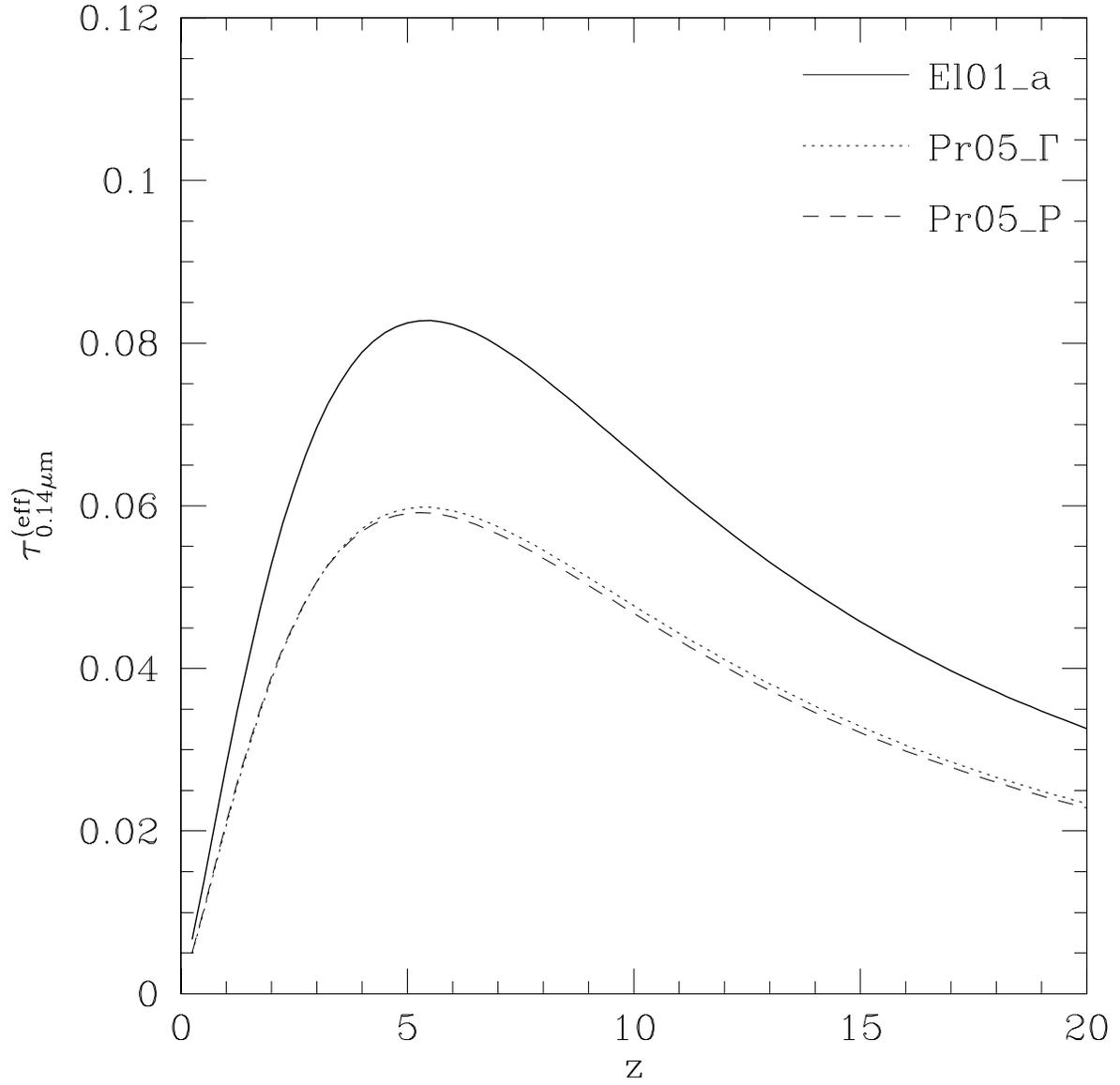} \caption{Effective optical depth at an
  emitted wavelength $\lambda_e=0.14 \mu m$ for the two Pr05 models
  compared with the El01\_a model.}\label{fig:pr}
\end{figure}
%%%%%%%%%%%%%%%%%%%%%%%%%%%%%%%%%%%%%%%%%%%%%%%%
\clearpage

%%%%%%%%%%%%%%%%%%%%%%%%%%%%%%%%%%%%%%%%%%%%%
\begin{figure}
  \plotone{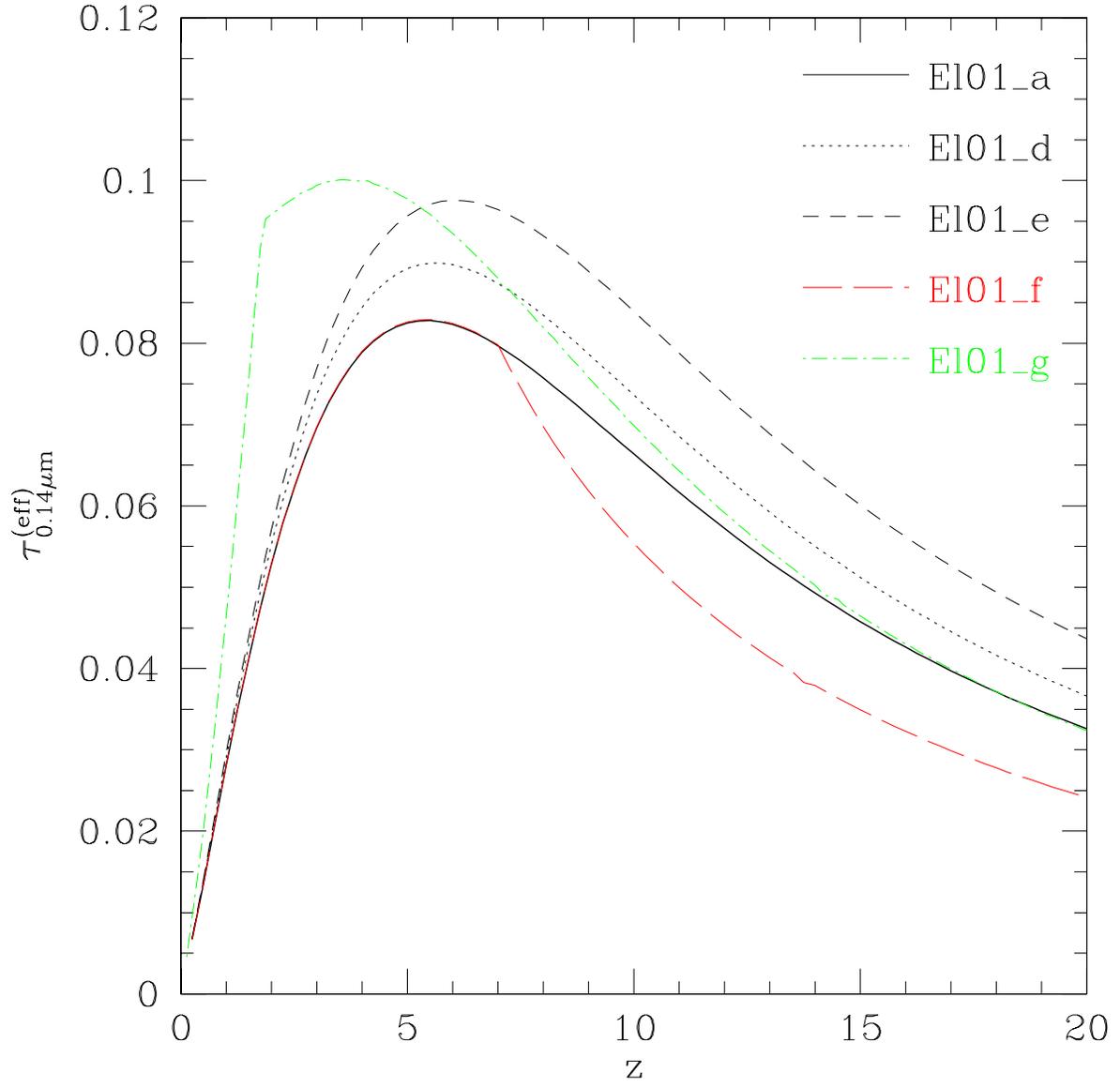} \caption{Effective optical depth at an emitted
  wavelength $\lambda_e=0.14 \mu m$ for different
  models.}\label{fig:comparison}
\end{figure}
%%%%%%%%%%%%%%%%%%%%%%%%%%%%%%%%%%%%%%%%%%%%%%%%

\clearpage

%%%%%%%%%%%%%%%%%%%%%%%%%%%%%%%%%%%%%%%%%%%%%
\begin{figure}
  \plotone{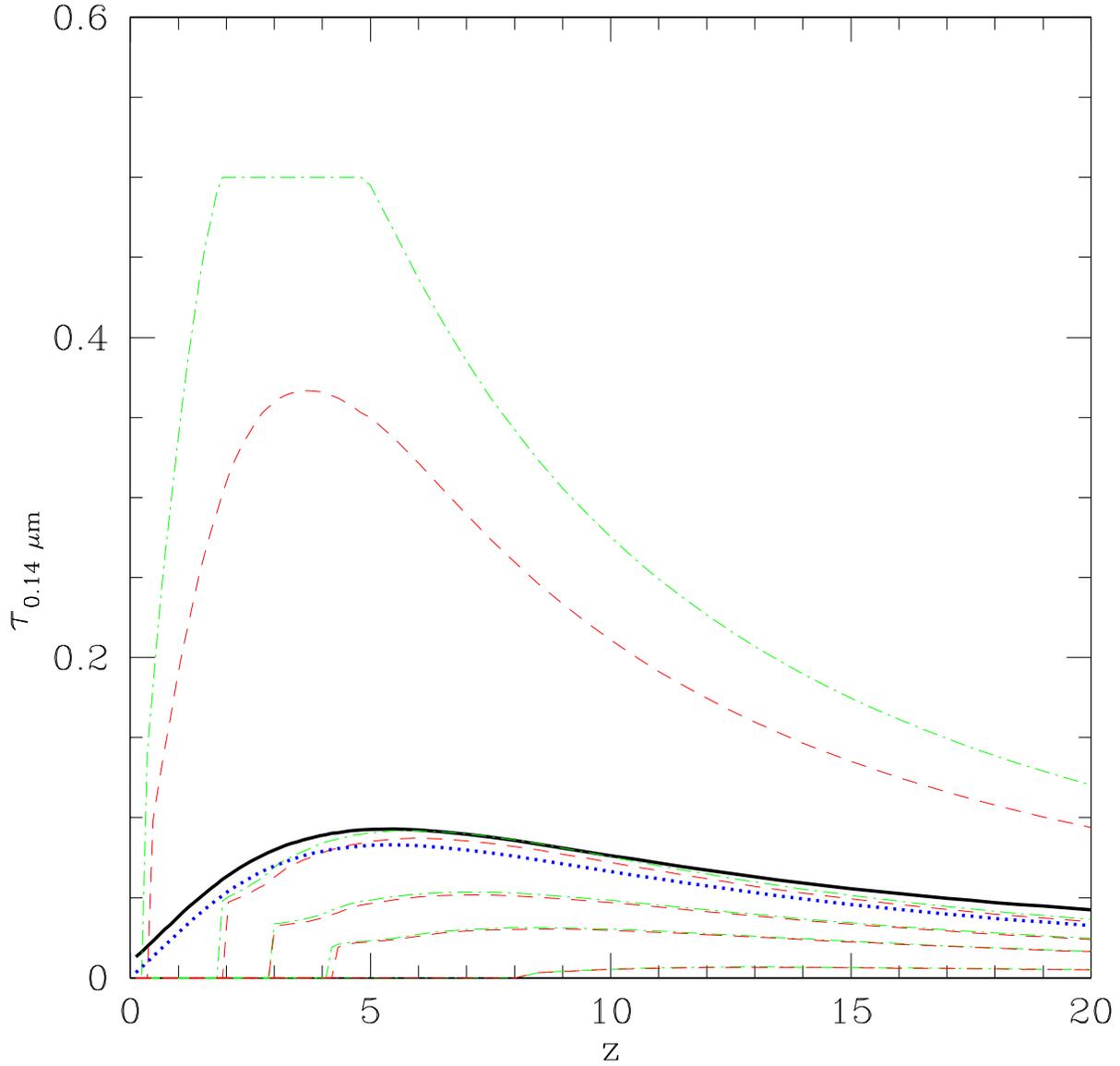} \caption{Optical depth vs. redshift for the
  model El01\_h compared with the model El01\_a. Effective optical
  depth is the bold black line for El01\_h and a bold dotted line for
  El01\_a. The green dashed-dotted lines represent for El01\_h,
  starting from the top, the upper $95\%$ contour in the distribution
  of optical depth, the upper $68\%$, the median, the lower $32\%$ and
  the lower $5\%$. The red dashed lines represent the same quantities
  for El01\_a.}\label{fig:el01k}
\end{figure}
%%%%%%%%%%%%%%%%%%%%%%%%%%%%%%%%%%%%%%%%%%%%%%%%

\clearpage

%%%%%%%%%%%%%%%%%%%%%%%%%%%%%%%%%%%%%%%%%%%%%
\begin{figure}
  \plotone{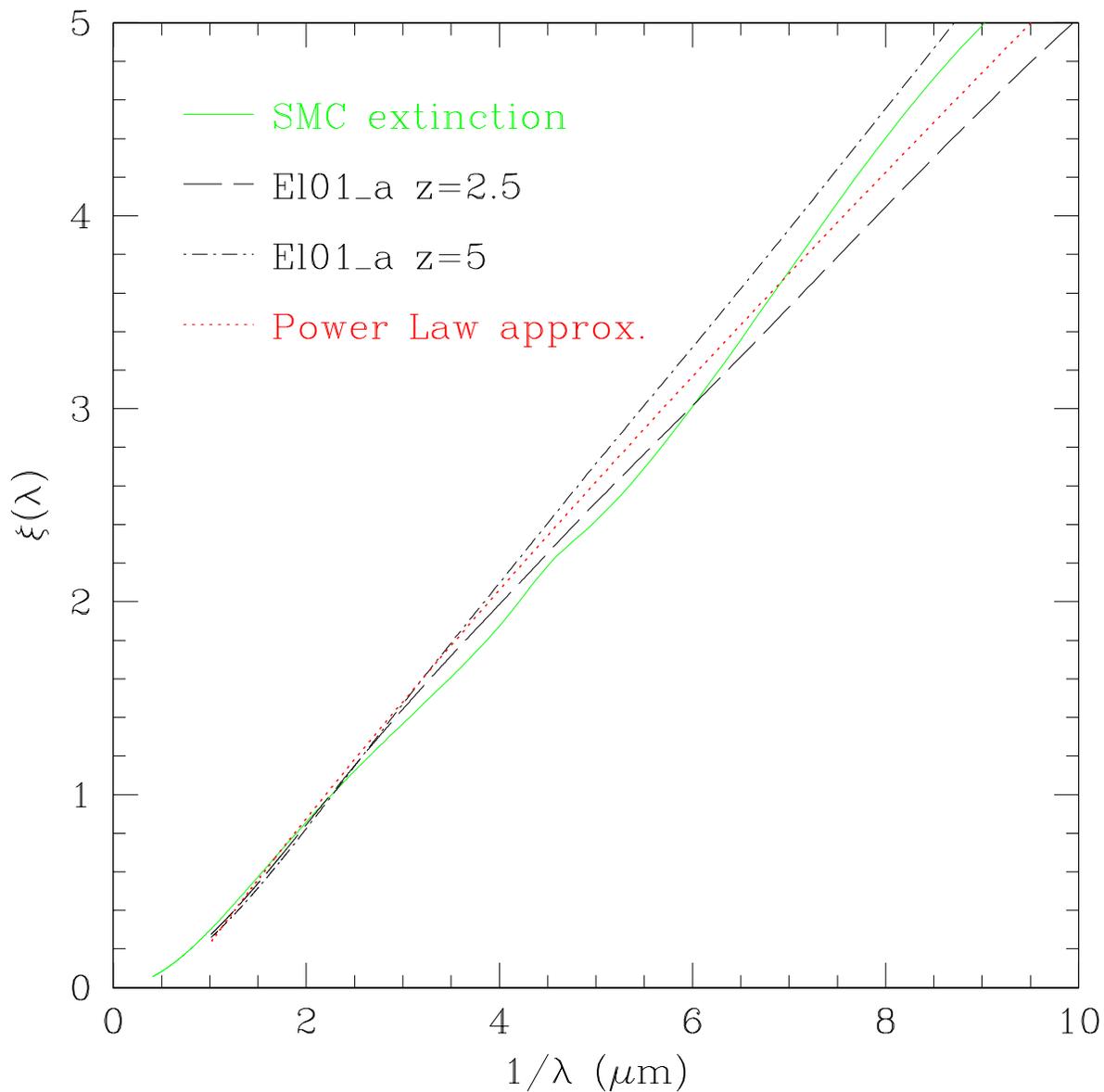} \caption{Effective mean extinction curve
  $\langle \xi(\lambda) \rangle =
  ln(E[q(\lambda)]/ln(E[q(\lambda_B)]))$ for the El01\_a model at
  $z=2.5$ (dashed) and $z=5$, (dotted-dashed) compared with the input
  SMC extinction curve (solid). Our analytical fitting formula,
  Eq.~\ref{eq:xi} (dotted line), is shown for
  comparison.}\label{fig:xi}
\end{figure}
%%%%%%%%%%%%%%%%%%%%%%%%%%%%%%%%%%%%%%%%%%%%%%%%

\clearpage

%%%%%%%%%%%%%%%%%%%%%%%%%%%%%%%%%%
\begin{table}
\begin{center}
\caption{Adopted parameters\label{tab:model}}
\begin{tabular}{lcccccc}
\tableline\tableline
ID& $A$ & $N_{\gamma}$ & $\alpha_1$ & $N_{min}$ & $N_{max}$ & $\alpha_{\kappa}$ \\ 
\tableline
Pr05\_$\Gamma$ & $0.0715$ & $3.03$ & $1.8$ & $0.2$ & $\infty$ & $1$ \\
\tableline
ID& $A$ &&  $\alpha_2$ & $N_{min}$ & $N_{max}$ & $\alpha_{\kappa}$ \\ 
\tableline
Pr05\_P & $0.0715 $ && $2.2$ & $0.2$ & $10$ & $1$ \\
El01\_a & $0.0910 $ && $2.1$ & $0.2$ & $10$ & $1$ \\
El01\_b & $0.0910 $ && $2.1$ & $0.2$ & $10$ & $0.5$ \\
El01\_c & $0.0910 $ && $2.1$ & $0.2$ & $10$ & $0.75$ \\
El01\_d & $0.0910 $ && $2.1$ & $0.2$ & $20$ & $1$ \\
El01\_e & $0.0910 $ && $2.1$ & $0.2$ & $100$ & $1$ \\
El01\_f & $0.0910 $ && $2.1$ & $0.2$ & $10$ & $\theta(7-z)$ \\
El01\_g & $0.0910 $ && $2.1$ & $0.2$ & $10$ & 1+{\small{`DLAbricks'}} \\
El01\_h & $0.0910 $ && $2.1$ & $0.2$ & $10$ & 1+{\small{`disks'}} \\
\tableline
\end{tabular}
%% Any table notes must follow the \end{tabular} command.
\tablecomments{Summary table with the parameters used to compute dust absorption.}
\end{center}
\end{table}

\end{document}